\documentclass[10pt, conference, compsocconf]{IEEEtran}
\ifCLASSINFOpdf
\else
\fi
\begin{document}
%
% paper title
% can use linebreaks \\ within to get better formatting as desired
\title{Cloud Services Brokerage: A Survey and Research Roadmap}

% author names and affiliations
% use a multiple column layout for up to three different
% affiliations
\author{\IEEEauthorblockN{Adam Barker, Blesson Varghese and Long Thai}
\IEEEauthorblockA{School of Computer Science \\
University of St Andrews, St Andrews, UK.\\ Email: \{adam.barker, varghese, ltt2 \}@st-andrews.ac.uk}
}

% conference papers do not typically use \thanks and this command
% is locked out in conference mode. If really needed, such as for
% the acknowledgment of grants, issue a \IEEEoverridecommandlockouts
% after \documentclass

% for over three affiliations, or if they all won't fit within the width
% of the page, use this alternative format:
% 
%\author{\IEEEauthorblockN{Michael Shell\IEEEauthorrefmark{1},
%Homer Simpson\IEEEauthorrefmark{2},
%James Kirk\IEEEauthorrefmark{3}, 
%Montgomery Scott\IEEEauthorrefmark{3} and
%Eldon Tyrell\IEEEauthorrefmark{4}}
%\IEEEauthorblockA{\IEEEauthorrefmark{1}School of Electrical and Computer Engineering\\
%Georgia Institute of Technology,
%Atlanta, Georgia 30332--0250\\ Email: see http://www.michaelshell.org/contact.html}
%\IEEEauthorblockA{\IEEEauthorrefmark{2}Twentieth Century Fox, Springfield, USA\\
%Email: homer@thesimpsons.com}
%\IEEEauthorblockA{\IEEEauthorrefmark{3}Starfleet Academy, San Francisco, California 96678-2391\\
%Telephone: (800) 555--1212, Fax: (888) 555--1212}
%\IEEEauthorblockA{\IEEEauthorrefmark{4}Tyrell Inc., 123 Replicant Street, Los Angeles, California 90210--4321}}

% use for special paper notices
%\IEEEspecialpapernotice{(Invited Paper)}

% make the title area
\maketitle

\begin{abstract}
A Cloud Services Brokerage (CSB) acts as an intermediary between cloud service providers (e.g., Amazon and Google) and cloud service end users, providing a number of value adding services. CSBs as a research topic are in there infancy. The goal of this paper is to provide a concise survey of existing CSB technologies in a variety of areas and highlight a roadmap, which details five future opportunities for research. 

\end{abstract}
% IEEEtran.cls defaults to using nonbold math in the Abstract.
% This preserves the distinction between vectors and scalars. However,
% if the conference you are submitting to favors bold math in the abstract,
% then you can use LaTeX's standard command \boldmath at the very start
% of the abstract to achieve this. Many IEEE journals/conferences frown on
% math in the abstract anyway.

% no keywords

% For peer review papers, you can put extra information on the cover
% page as needed:
% \ifCLASSOPTIONpeerreview
% \begin{center} \bfseries EDICS Category: 3-BBND \end{center}
% \fi
%
% For peerreview papers, this IEEEtran command inserts a page break and
% creates the second title. It will be ignored for other modes.
\IEEEpeerreviewmaketitle

\section{Introduction}

The public cloud computing \cite{hotcloud} market is crowded with competitors and choices, each offering their own set of heterogeneous services (e.g., storage, elastic-compute) and configuration options, covering for example: Virtual Machine (VM) instance type (e.g., high-memory, high-CPU etc.), data centre region, availability zone etc. Often it is the case that users deploy applications onto cloud infrastructure on an ad hoc basis, without understanding which providers and instance types can meet user-driven Service Level Objectives (SLOs): individual measurable performance metrics, which cover Quality of Service (QoS) properties such as availability, throughput and response time.

A \emph{Cloud Services Brokerage (CSB)} acts as an intermediary between cloud service providers (e.g., Amazon and Google) and cloud service end users.  Brokerages utilise several types of brokers and platforms to enhance service delivery, and ultimately service value \cite{gartner}. CSBs as a research topic are in there infancy:  although there have been numerous research projects which have attempted to develop individual components for a CSB, there have (to date) been no successful integrated frameworks which provide an end-to-end solution. The goal of this paper is to provide a concise survey of existing CSB technologies in a variety of areas and highlight a roadmap, which details five key opportunities for  research.

%Gartner define three main categories of service that brokers can provide: \emph{intermediation, aggregation} and \emph{arbitrage}. An \emph{intermediation} broker adds value on top of a given service, enhancing a specific capability. An \emph{aggregation} broker brings multiple services together, potentially repackaging its use for an end-user. An \emph{arbitrage} broker provides opportunistic choices for end-users, potentially buying and selling resources in different markets. 

%Definition of a CSB and why it's useful: 
%Objective of the research paper

\section{Survey}

Our survey of existing CSB technologies has been broken down into four key categories. Firstly \emph{CSBs for performance}, addresses issues of cloud performance comparison and prediction. \emph{CSBs for application migration}, discusses decision support systems that help guide a developer when making a decision about whether to migrate to the cloud.  \emph{Theoretical models for CSBs}, discusses purely theoretical and mathematical techniques. Finally \emph{data for CSBs} presents an overview of providers that supply raw data or metrics, which can be used as input to drive a CSB.  

\subsection{CSBs for Performance} 

\emph{STRATOS}~\cite{stratos} is a cloud broker service, which focuses on solving a Resource Acquisition Decision (RAD) problem, involving the selection of \emph{n} resources from \emph{m} cloud providers. STRATOS requires two sets of inputs in order to solve the RAD problem. The first is the application topology to be deployed on the cloud, this is specified through a Topology Descriptor File (TDF). The second is a set of developer objectives: measurable constraints, requirements and preferences. The Service Measurement Index (SMI)~\cite{SMI} is an approach to facilitate the comparison of cloud services. SMI consists of seven categories relating to accountability, agility, assurance, financial, performance, security, privacy and usability, with each category being further refined to a set of attributes, expressed as a set of Key Performance Indicators (KPIs). STRATOS utilises KMIs in order express a developer's objectives. STRATOS then solves the RAD problem at runtime through a multi-criteria optimisation problem utilising the desired configuration (defined through a TDF), a set of objectives (defined through KMIs) and data from third-party services such as CloudHarmony~\cite{cloudharmony}.

%A TDF (represented through XML) consists of a set of clusters, where each cluster is comprised as a set of nodes. A node is simply a representation of a VM instance type specifying characteristics such as CPU, RAM, disk as well as more abstract metrics such as performance and security. 

\emph{OPTIMIS}~\cite{optimis} is a multi-cloud toolkit aimed at infrastructure providers, service providers and application developers. It provides a suite of functionalities aimed at optimising IaaS clouds, covering the full lifecycle from service construction, cloud deployment and operation. OPTIMIS focuses on key non-functional concerns, regulatory and legislative constraints. A broker component is contained within the OPTIMIS toolkit, which allows alternative cloud configurations to be directly compared in terms of business efficiency.

\emph{CloudCmp}~\cite{cloudcmp} focuses on measuring the performance of a range of cloud services including: elastic compute clusters, persistent storage, intra-cloud networking and wide-area networking. Each broad service area is evaluated through a series of individual benchmarks and metrics. \emph{Computation metrics} are evaluated through Java-based benchmarks from SPECjvm2008, covering the completion time of the benchmark, cost-per-benchmark, and scaling latency: time to start new VMs. \emph{Storage metrics} address the response time, throughput, time to consistency and cost per operation. Finally \emph{networking metrics} are evaluated through standard tools such as iperf and ping. 

\emph{Conductor}~\cite{Wieder:2012:NSDI} is a framework which focuses on optimising the execution of a MapReduce application on the cloud by utilising resources based on a user's goals, such as minimising cost or completion time. It also supports dynamic monitoring and utilisation in order to handle prediction error or system failure on runtime.

\emph{Matrix}~\cite{Chiang:2014:ICAC} is a recent research project that is able to predict the performance of a given application on different hardware types based on the behaviour similarity between an application and previously sampled ones. However, it focused on either predicting the performance or selecting the most cost effective instances instead of providing options based on a user's requirements.

\emph{CloudBench}~\cite{benchmarking} is a benchmarking methodology and prototype framework in which a user provides a set of four weights that indicate how important each of the following groups: memory, processor, computation and  storage are to the application that needs to be executed on the cloud. The weights along with cloud benchmarking data are used to generate a ranking of VMs that can maximise performance of the application.

\subsection{CSBs for Application Migration}

\emph{CloudProphet}~\cite{cloudprophet} is a tool focused on application migration to the cloud. CloudProphet traces (records) the workload of the application when running locally, and replays the same workload in the target cloud(s) for prediction. Once the workload is replayed in multiple target clouds a decision can be made about which cloud to migrate the application; typically this can be a costly and time consuming process. 

\emph{The Cloud Adoption Toolkit}~\cite{Khajeh-Hosseini} addresses a number of challenges that decision makers face when assessing the feasibility of the adoption of cloud computing in their organisations. The most mature component focuses on cost prediction, and allows an organisation to model their application requirements over time and predict migration and future costs across multiple cloud providers. 

\emph{RightScale}~\cite{rightscale} offers a number of services, which allows users to deploy and manage applications across multiple cloud providers. Basic brokering mechanisms are provided through an alert-action mechanism. VMs managed by RightScale have a number of pre-defined hooks, which send data back to the RightScale console, this allows an alert to trigger an associated action, e.g, a scaling policy \cite{SPE_survey}.

\subsection{Theoretical Models for CSBs}

\emph{SMICloud}~\cite{Garg:2013} is a decision support tool, which allows users to evaluate and rank clouds based on a user's Quality of Service (QoS) requirements. SMICloud is based upon the Service Measurement Index (SMI)~\cite{SMI} proposed by the Cloud Service Measurement Index Consortium, also used by STRATOS. SMICloud takes into account functional and non-functional requirements and covers: service response time, sustainability, data centre performance per energy (covering carbon emissions), suitability, accuracy, transparency, interoperability, availability, reliability, stability, cost, adaptability, elasticity, usability, throughput and scalability. An Analytical Hierarchical Process (AHP) based ranking mechanism is proposed as a way of evaluating user defined QoS requirements against cloud providers. 

Authors of~\cite{cloud_selection} \emph{mathematically model} the cloud service selection problem and present a multi-criteria methodology, which focuses on instance selection for performance.

\subsection{Data for CSBs}

\emph{CloudHarmony}~\cite{cloudharmony} is a commercial product, which provides four core services so that users can compare a large set of commercial cloud providers. \emph{CloudSquare} allows users to directly compare the features of cloud providers, e.g., providers offering compute at 100\% availability in Europe. \emph{CloudScores} offers a set of benchmarking data for popular cloud providers; currently SPECint, SPECfp, Geekbench, Geekbench Multicore, Unixbench and Unixbench Multicore data are available.  \emph{CloudMatch} provides realtime metrics for cloud providers, e.g., networking latency for all providers, upload speed etc. Finally \emph{CloudReports} offers many of the previous services packaged into either quarterly or monthly reports. CloudHarmony have a number of Web service (both REST and SOAP) APIs available, allowing application developers to utilise these metrics for cloud brokers. 

A recent \emph{EPSRC project}~\cite{gillam2013fair} discussed a set of ``fair" benchmarks for cloud computing systems. The report provides a comprehensive survey of existing benchmarking techniques and presents the results of performance benchmarking covering: Memory IO, CPU, Disk IO, Application and Network for Amazon, Rackspace, IBM and a private cloud installation of OpenStack. The research was motivated towards performance-based pricing schemes, e.g., paying more for better, or more reliable performance.

The \emph{Yahoo! Cloud Serving Benchmark (YCSB)}~\cite{YCSB} is an open-source framework for evaluating the retrieval and maintenance capabilities of NoSQL database systems.

\section{Research Roadmap}

This Section discusses a set of research opportunities in the form of a roadmap for cloud brokerage. Each opportunity forms an initial step in our idealised broker application. We are proposing further research in the following areas: capturing user-driven requirements, application specific brokering, lightweight cloud benchmarking, dynamic cloud management and a unified Broker as a Service (BaaS) framework. 

\subsection{Capturing High-Level User Requirements}
\label{roadmap1}

Most of the current research efforts on building cloud brokerage systems has focused on a user providing a low-level specification of the hardware required, e.g., two instances with 4Gb of memory, with 4 vCPUs. It is very difficult for users to translate these low-level hardware specifications into the overall desired performance required by an application. Research needs to address how users specify high-level Service Level Objectives (SLOs), which specify a desirable goal for a deployment, examples of goals include: end-to-end latency, throughput, program execution time, high availability etc. \cite{metrics} provides an interesting discussion of ``good'' performance metrics for cloud services brokers. 

Research could take the form of a high level policy language and associated toolchain, which translates user defined goals (e.g., maximum end-to-end latency) into a suitable deployment consisting of a set of instance types and geographical regions across multiple cloud providers. Very recent research has also hinted at this problem under the name Software Defined Service Level Agreements (SLAs) \cite{slas}.

\subsection{Application Specific Brokering}
\label{roadmap2}
%Long: 

Each user wanting to deploy an application on the cloud has a different (and potentially unique) set of requirements. For example, while a developer of a Hadoop application requires that a certain number of tasks must be completed in a given time frame, an owner of a web server wants her application to be able to respond to any request within a certain amount of time, regardless of the current workload. Similarly, an organisation with a high budget is able to pay a lot of money to minimise its application's execution runtime, while a small start-up might prefer a cheaper option, which takes longer to complete. As a result, it is necessary to have a cloud broker service, which takes into account an application's behaviour in order to suggest the best options based on a user's requirements. As all applications behave differently, the decision made by a broker regarding one application cannot necessarily be applied to another.

Developing a cloud broker which makes decisions based on application specific characteristics is challenging as it requires not only complete knowledge of both the application's behaviour and cloud resource performance, but also requires a method, which can find the connection between these two data sets in order to make an accurate prediction. We advocate further research upon application modelling and workload classification, capture and prediction. Furthermore, instead of just maximising performance or minimising the execution cost of an application, a broker must be able to enforce a user-defined QoS property based on the application model and predicted workload. CloudProphet \cite{cloudprophet} and Matrix \cite{Chiang:2014:ICAC} have taken some first steps towards this goal.

%We advocate further research upon application modelling, workload classification capture and prediction. CloudProphet\cite{cloudprophet} as discussed in Section [] has taken some first steps towards this goal. 

\subsection{Lightweight Cloud Benchmarking}
\label{roadmap3}

There are a number of efforts to benchmark cloud resources (for example, \cite{benchmarking,cloudrankd,Iosup:2011:PAC}). Either standard tools or custom-built tools are employed to generate numerous metrics that reveal a number of attributes of cloud resources. Most benchmarking methods are however heavy weight (time consuming) and cannot be used in real-time. Therefore the data generated has to be used as historic data for predicting the performance of the resource. To be able to make accurate predictions it would be valuable to have historical benchmarking data, which can be used in conjunction with live monitoring and benchmarks. 

We propose, that a public (potentially crowdsourced) benchmarking repository be made available for the cloud community, which captures metrics across a range of cloud providers and instance types. By providing applications with historical data across different cloud providers it is more likely that applications can accurately predict and schedule applications given user defined Service Level Objectives (SLOs). 

In addition we propose lightweight benchmarking methods to be developed, such that benchmarks can be deployed and executed in near realtime; this is usually in contrast with most benchmarking methods, which as discussed above are time consuming processes. Cloud WorkBench \cite{cloudworkbench} is a first step towards achieving this objective and builds upon the Infrastructure-as-Code concept.

%Service Level Agreements (SLAs) 

%on the cloud in terms of accurate Service Level Agreements (SLAs)Hence, consistent and historic benchmarking data can be obtained for the broker for executing a different application. The challenge however is going to be the willingness of users and managing the size of the rapidly growing repository. 

%Instead of users uploading benchmarking data into the repository, we suggest that users of applications that make use of a unified BaaS framework API be willing to allow the broker to store the benchmarked data of the application. Hence, consistent and historic benchmarking data can be obtained for the broker for executing a different application. The challenge however is going to be the willingness of users and managing the size of the rapidly growing repository. 

\subsection{Dynamic Cloud Management}
\label{roadmap4}

During the execution of an application, the broker service must be able to dynamically reschedule an application on the cloud resources to ensure that the high-level requirements of a user are always met, even when the application is still running. For example, assume the user wants to ensure that the end-to-end latency of an application is less than 100 milliseconds, and the network latencies during the execution at the given provider(s) are increasing. The supplied latency user requirement will not be satisfied in the long-term and the broker needs to dynamically monitor and manage the deployment; this might involve: migrating the application onto different resource types, different providers or regions, adding or removing VMs etc. While significant efforts have been made in this direction, for example \cite{monitoring1,monitoring2}, they still need to be integrated with CSB services in a meaningful way. 

\subsection{Broker-as-a-Service (BaaS)}
\label{roadmap5}

%The ideal use of a broker is to not only take the requirements of a user and determine the best cloud provider or resources for an application, but also map the requirements efficiently in a fine-grain manner onto the resources provided by multiple providers. For example, if resources are being offered from a number of cloud providers such as Google Compute Engine, Amazon and Rackspace it must be possible for the broker to map the requirements such that the combination of resources from all three providers is leveraged if it can maximise performance and reduce overall running costs. To develop such a brokering service, benchmarking data needs to be obtained from cloud resources of multiple providers. Usually benchmarking data that is available are low-level metrics related to the performance of the resources. The challenge, however, is in (i) the availability of benchmarking data in a consistent manner (also considered in Section \ref{roadmap3}, and (ii) the interpretation and aggregation of low level metrics to the requirements (addressed in Section \ref{roadmap2}. 

As a future direction, we are proposing a publicly available cloud Broker-as-a-Service (BaaS) framework, which can provide the following across multi-cloud environments: (i) capture high-level user requirements and translate these requirements into concrete resources (instance types, regions etc.), discussed in Section \ref{roadmap1}; (ii) provide functionalities for application specific brokering, discussed in Section \ref{roadmap2}; (iii) facilitate lightweight benchmarking and utilise benchmarking data through a historical repository, discussed in Section \ref{roadmap3}; (iv) and allow dynamic reconfiguration of resources, discussed in Section \ref{roadmap4}. The BaaS framework can be made available through a publicly accessible Web-based (e.g., REST) API that can automatically provide a suite of brokering services for applications.

\section{Conclusion}

This paper has reviewed the current state-of-the-art in Cloud Services Brokerage (CSBs), an important emerging topic in the cloud computing research community. After reviewing the state-of-the-art this paper then presented a research roadmap through an idealised broker, which requires new research contributions in five key areas: capturing and translating high-level user requirements, application specific brokering, new lightweight benchmarking techniques, dynamic cloud management, all of which should be integrated into a unified Broker as a Service (BaaS) framework.

\bibliographystyle{abbrv} \bibliography{../broker}

% that's all folks
\end{document}